\documentclass[12pt]{article}
\usepackage{pdfpages}
\pdfoutput=1
\def\ni{\noindent}
\def\siml{\underline{\sim}}

\def\np{\newpage}
\def\part{\partial}
\def\dint{{\rlap{=}\! \int}}
\def\dsint{{\rlap{=}\int}}
\def\chtw{\chi_{2}^2}
\def\ynu{y_{\nu}}
\def\phih{\hat{\phi}}
\def\sgh{\hat{\sigma}}
\def\yk{y^{(k)}}
\def\Ntw{[{N \over 2}]}
\def\Ab{{\bf{A}}}
\def\Eb{{\bf{E}}}
\def\Vbr{\bar{V}}
\def\jb{\bar{j}}
\def\wb{\bar{W}}
\def\wbr{\bar{W}}
\def\ybr{\bar{y}}

\def\Ebf{{\bf{E}}}

\def\Sbr{\bar{S}}
\def\Sh{\hat{S}}
\def\Shb{\overline{\hat{S}}}
\def\Shk{\hat{S}^{(k)}}
\def\Shnu{\hat{S}_{\nu}}

\def\SI{ \hat{S}_{(\bar{i})} }
\def\SJ{ \hat{S}_{(\bar{j})} }
\def\SD{ \hat{S}_{(\bar{\cdot})} }
\def\Vt{\tilde{V}}
\def\cd{\cdot}
\def\kv{\vec{k}}
\def\xv{\vec{x}}

\def\th{\theta}
\def\om{\omega}
\def\om{\omega}

\begin{document}
\begin{center}
{\bf Spectral  Estimation of Plasma \\
Fluctuations I: Comparison of Methods} \\
\end{center}
\begin{center}
{\bf Kurt S. Riedel and Alexander Sidorenko,\\
Courant Institute of Mathematical Sciences \\
New York University \\
New York, New York 10012-1185 \\
\ \\
David J. Thomson \\
AT\&T Bell Laboratories \\
Murray Hill, NJ 07974-0636  \\
\  \\
}\end{center}

\begin{abstract}
The relative root mean squared errors (RMSE) of nonparametric 
methods for spectral estimation is compared for 
microwave scattering data of plasma
fluctuations. These methods reduce
the variance of the periodogram estimate by averaging the spectrum
over a frequency bandwidth. As the bandwidth increases, the variance decreases,
but the bias error increases. 
The plasma spectra vary by over four orders
of magnitude, and therefore, using a spectral window is necessary. 
We compare the smoothed tapered
periodogram with the adaptive multiple taper methods and hybrid methods.
We find that a hybrid method, which uses four orthogonal tapers and then
applies a kernel smoother, performs best. For 300 point data segments, even an
optimized smoothed tapered periodogram has a  24 \% larger
relative RMSE than the hybrid method. We  present
two new adaptive multitaper weightings which outperform 
Thomson's original adaptive weighting.  
\end{abstract}

\ \\


\newpage
\noindent
{\bf I. Introduction}

The measurement and analysis of plasma fluctuations have become
increasingly important as the magnetic fusion community attempts to
understand and cure plasma turbulence. Two excellent reviews
summarize the experimental work on plasma fluctuations and anomalous
tokamak transport.$^{1-2}$ To understand the precise nature of the drift
wave turbulence, we hope to determine the experimental dispersion
relation, the rollover points, $\omega_R$, of the fluctuation spectrum
(the value of $\omega$ and $\vec{k}$ where the fluctuations are the largest),
and the spectral decay exponents.

In this article, 
our goal is to more efficiently and accurately determine the spectral
distribution of plasma fluctuations for a given length digitalized time
series. 
We compare a number of different methods$^{3-10}$  to estimate
the spectral density of fluctuations using actual fusion data.
We concentrate on time stationary fluctuations. In a successor article, we
present a number of applications of multitaper spectral analysis$^{3-7}$ 
to nonstationary times series$^{11,12}$.

We analyze time series data from the Tokamak Fusion Test Reactor$^{13}$
(TFTR) microwave scattering diagnostic$^{14-17}$. 
Because the TFTR spectrum varies over 
four orders of magnitude, the standard techniques$^{8-10}$ 
of spectral analysis do not necessarily perform well. 
Therefore, we compare the estimation methods on data which is typical
of forward scattering experiments.  
All of the spectral estimation methods in our study
yield nearly identical estimates
for the entire 45,000 point data segment of interest.
When the plasma fluctuations are nonstationary, much shorter
time segments must be used. Therefore, we compare the relative accuracy
of several spectral estimations methods on data segments of 300-3000
points. {\it For these short sample lengths, we show that the advanced 
analysis techniques significantly outperform the conventional estimates.}

All of the methods essentially average over some frequency bandwidth
to reduce the variance. As the bandwidth is increased, more degrees of
freedom are used. If the bandwidth is too large, the spectral estimate will
be artificially broadened, and a bias error will result. Thus selecting
the ``correct'' bandwidth is crucial, and we choose the optimal global
bandwidth for each method separately.

In Section II, we describe the physics of the TFTR microwave scattering 
experiment. 
In Section III, we review the standard approach to 
estimating spectral densities. 
In Section IV, 
we describe and apply a family of advanced statistical
techniques for estimating spectral densities: 
multitaper spectral methods$^{3-7}$. 
Multiple taper methods have been successfully applied to
a number of different problems in geophysics. 
Sections III and IV present some of the results of the
comparison described in Sec. VI.
In Section V and Appendix C, we describe the jackknife procedure$^{18-20}$ 
for nonparametric variance estimation.

In Section VI, we compare the accuracy of the various spectral
estimation techniques using the  TFTR data. We examine the relative
accuracy of the smoothed  periodogram and the multiple taper analysis
for 300--3000 point data subsegments.
We show that both methods yield the nearly identical estimates 
for long data segments and that multitaper converges more rapidly.  
Unexpectedly, we find that {\it a hybrid method where the multitaper spectral
estimate is then smoothed performs appreciably better} than either the
smoothed periodogram or the  ``pure'' multitaper.
We summarize our results on the estimation of TFTR spectra in Sec.~VII.
In Appendix A, we summarize the properties of the Slepian
tapers$^{21}$ which form the bases for multitaper analysis. In Appendix B,
we present two new adaptive weightings for  multitaper estimation:
sequential deselection and the minimal expected loss weighting.
Appendix D gives the details of our  empirical RMSE comparison. 

\ \\
\noindent
{\bf II. TFTR microwave scattering dataset} 
\vspace{.2in}

We now describe the TFTR data and the physics of the underlying  
fluctuations. The microwave transmitter launches a 60 GHz plasma wave 
linearly 
polarized in the extraordinary mode below the electron cyclotron 
frequency$^{14,15}$.
The 112 GHz extraordinary mode plasma wave propagates from the top 
of the plasma to the core plasma, where the plasma density 
fluctuations scatter the incoming wave$^{22-25}$. 
The scattered wave is measured by a receiver 
located near the bottom of the vacuum vessel so that
$|\kv_{scat}- \kv_{inc}| \siml 3$ cm and $\kv$ is parallel to the 
poloidal magnetic field at $\frac{r}{a} \siml .3$.
The scattered signal is proportional to the density fluctuations. 

The plasma waves are believed to consist of an ensemble of 
drift waves$^{1,2,25}$.
For electron drift waves, the characteristic frequency is the electron
drift frequency $\omega \sim \omega_{D_e} \equiv dn_e /dr$ and the
characteristic distance is $\rho_S \siml \rho_i \sqrt{T_e /T_i} $, where
$\rho_i$ is the ion-cyclotron radius. Since the drift waves are localized
near the Doppler-shifted resonance surface, the drift wave dispersion
relation is approximately $\omega \sim V_{D_e} k_{\theta}$. Including the 
zeroth order Larmor radius effects modifies the dispersion relation
to
$\omega \sim v_{D_e} k_{\perp} /(1 + k_{\perp}^2 \rho_S^2 )$,
where $v_{D_e} \equiv \frac{T_e}{eB}\frac{d \ln(n_e)}{dr}$. 
From mixing length theory, we expect the saturated fluctuation amplitudes
to be approximately
$$
{\delta n \over n } \sim {e \delta \Phi \over \kappa T_e} \sim
{1 \over k_{\perp} L_n}
\ ,$$         
where $L_n$ is the equilibrium density gradient scale length.

For TFTR discharge number 50616, the plasma parameters are
$B_t = 4$ Tesla, $I_p = 1.2$MA, $\overline{n} = 3.6 \times 10^{15}$ cm$^{-3}$. 
The central electron temperature is approximately 2 Kev 
and the central ion temperature is approximately 1 Kev. 
In the scattering volume, the local plasma parameters are
$\rho_i = 0.08$ cm, $\rho_S$ = 0.11 cm, $k_{\theta} \siml 3$ cm$^{-1}$,
$V_{D_E} = 4.9 \times 10^4 $m/sec, $\omega_{D_e}/2\pi = 23$ kHz.

In most drift wave theories, both short and long wavelength drift waves
are stable, and only moderate wave length drift waves are unstable. 
We expect the largest amplitude fluctuations to occur 
at $k \rho_S \sim .1-.5$. Typically, 
$\omega_{measured}\siml (2-4 \omega^*_e$, possibly due to toroidal plasma
rotation. 

The time series begins 5.0 sec. into the discharge, and
is totally contained in the Ohmic phase.
We redefine time equals zero $(t \equiv 0)$ as the start of our series. 
A macroscopic sawtooth oscillation occurs at 
3.86 millisec into the dataset.
During this time, the plasma is strongly nonstationary and 
the analysis requires the methods of our successor article$^{11}$.
The scattering volume lies just outside the sawtooth mixing radius.

Our data consists of 65,500 time samples 
with a uniform sampling rate 5 MHz over the time interval.
Thus the fluctuations are recorded over a tenth of a second time interval.
The data has been low-pass filtered with an anti-aliasing filter
with a filter halfwidth of 2.5 MHz.
The spectral peak at 1 MHz corresponds to the receiver intermediate frequency
(IF), and is caused by wall and waveguide reflections.
The broadening of the 1 MHz peak is believed to
be due to intense edge fluctuations$^{16,17}.$ 

In this article, we concentrate on obtaining the best possible estimates 
of the fluctuation spectrum  during the stationary part of the 
discharge. Thus we consider only the 45,000 datapoints beginning .24 millisec
after the sawtooth. 


\newpage
\noindent
{\bf III. Single Taper Spectral Analysis}
\vspace{.2in}

{\sc A. Integral Formulation of Stationary Processes}

\ \\
We  consider a stationary stochastic process with no
deterministic spectral lines. We are given $N$ discrete measurements,
$\{ x_{-\Ntw},\ldots ,x_0,x_1 , \ldots , x_{\Ntw} \}$,
of a realization of the stochastic process. 
For convenience, we assume that the data length is an odd integer, and
define $\Ntw \equiv ({N-1 \over 2})$. We index the data such that
the midpoint of the data is measured at time equal to zero.
The Nyquist frequency is ${1\over2 \Delta t}$,
and the Rayleigh resolution frequency is ${1 \over N \Delta t}$.
We define the time-frequency bandwidth, $\wb\equiv WN\Delta t $, where
$W$ is the frequency bandwidth.
We normalize the time interval, $\Delta t$, between measurements to unity. 

The Cramer representation of a stationary stochastic process$^{8-10}$ is
$$
x_n = \int_{-1/2}^{1/2} e^{2 \pi ifn} dZ(f)  \ , \eqno (1)
$$
where $dZ$ is a random measure. Stationarity implies that 
$dZ$ has independent spectral increments, i.e. the values of
$dZ$ at different frequencies, $f$ and $g$, are uncorrelated.
We assume that the spectral measure is absolutely continuous with a spectral
density $S(f)$. Thus the covariance of $dZ$ satisfies
$$
\Ebf[dZ(f)d \bar{Z} (g)] = S(f) \delta (f-g) dfdg . \eqno (2)
$$
The measured time series, $\{ x_i \}$, has a discrete Fourier
transform, $y(f)$:
$$
y(f) = \sum_{n=-\Ntw}^{\Ntw} x_n e^{-2 \pi inf} \ , \eqno (3)
$$
and a corresponding inverse:
$$
x_n = \int_{-1/2}^{1/2} y(f) e^{+2 \pi inf} df \ . \eqno (4)
$$
Substituting Eq.~(1), we obtain the Cramer representation of the
discrete Fourier transform,
$$
y(f) = \int_{-1/2}^{1/2} \sum_{n=-\Ntw}^{\Ntw} e^{- 2 \pi in(f-f^{\prime})}
dZ(f^{\prime} )
= \int_{-1/2}^{1/2} {\sin (N \pi (f-f^{\prime})) \over
\sin ( \pi (f-f^{\prime}))}  
dZ(f^{\prime} ) 
\ \ .\eqno (5)
$$
Thus the discrete Fourier transform of the measured process is related
to the realization of the stochastic process by an integral equation of the
first kind. Similarly, the expectation of the discrete periodogram, 
$I(f) \equiv |y(f)|^2$, is related to
the spectral distribution, $S(f)$, via a convolution equation with the
Fej\'{e}r kernel:
$$
\Ebf[I(f)] \equiv \Ebf[|y(f)|^2 ] = \frac{1}{N}\int_{-1/2}^{1/2}
\left[ {\sin (N \pi (f-f^{\prime})) \over \sin ( \pi (f-f^{\prime}))}
\right]^2 \
S(f^{\prime} )df \ . \eqno (6)
$$
We note that {\it Eq.~(5) applies to the particular realization of the
stochastic process, $dZ(f)$, which has been observed, and that Eq.~(6)
applies to the expectation of the periodogram.}
In the next subsection, we show that the standard deviation
of the simple periodogram estimate, $|y(f)|^2$, is approximately equal to 
the spectrum itself$^{8-10}$. {\it Thus the variance of the periodogram
estimate is large, and does not converge with $N$.}

To estimate the bias of the simple periodogram estimate of the
spectrum, we make a Taylor series expansion of $S(f)$  about $f$:
$S(f + \tilde{f} ) \sim S(f) + S'(f)\tilde{f} +S''(f)\tilde{f}^2/2$
and substitute the expansion into Eq.~(6): 
$$
\Ebf[I(f)] \sim S(f) + {S''(f)\over 2}
\int_{-1/2}^{1/2} \frac{(f-f^{\prime})^2}{N}
\left[ {\sin (N \pi (f-f^{\prime})) \over \sin ( \pi (f-f^{\prime}))}
\right]^2 \
 df' \ . \eqno (7)
$$
The standard approach to reducing the bias of spectral estimate
is to apply a data taper, and thereby reduce the sidelobes of the
kernel. \vspace*{.2in}

\ \\
{\sc B. Brief Overview of  Spectral Estimation}
\ \\

The inverse spectral problem is to determine $S(f)$, given the
time series data:
$\{ x_{-\Ntw},\ldots ,x_0,x_1 , \ldots , x_{\Ntw} \}$.
The inverse problem can be viewed as a coupled set of integral
equations: Eq.~(5) relates $y(f)$ to the random process, $dZ(f)$,
and Eq.~(6) relates $E|y(f)|^2$ to the spectral density $S(f)$.

Our basic goal is to invert the Fej\'{e}r integral equation, Eq.~(6),
to determine $S(f)$. Equation~(6) is an integral equation of the
first type. The standard numerical method for the inversion
of integral equations of the first kind is to regularize the
equation by adding a small smoothness penalty function of the
form $\int [S''(f)]^2 df$. Spectral estimation contains essentially
three difficulties which require more specialized techniques.

First, the values of the estimated periodogram, $I(f)$, are correlated
in the frequency domain, and therefore, the estimation technique
needs to take account of this correlation to be effective.

Second, most numerical inversion techniques produce broader and
smoother spectral estimates than the true spectra. This bias is
especially disadvantageous when the spectrum has one or more narrow 
peaks and the shape of the spectrum is of primary interest. In turbulent 
plasma fluctuations, the exponent of the spectral rolloff helps
to determine the nature of the turbulent fluctuations.

Third, Eq.~(6) assumes the spectrum is purely incoherent and uses
only the amplitude of $y(f)$. Thus the phase information in $y(f)$
is ignored. 
When the spectrum contains coherent components such as spectral lines,
the phase information is crucial in estimating the coherent component.
Therefore, we attempt to invert Eq.~(5), 
and determine the realization of $dZ(f)$ before estimating $S(f)$.

\ \\
{\sc C. Bias and Variance of Single Taper Estimates}
\  \\


The standard single taper theory of nonparametric spectral analysis$^{8-10}$
multiplies the data by a taper, $\nu_n, \ n=-\Ntw, \ldots ,\Ntw$, prior
to performing the Fourier transform, in order to reduce the bias
from the sidelobes of the Fej\'{e}r kernel.
Thus the tapered transform is
$$
y_{\nu}(f) = \sum_{n=\Ntw}^{\Ntw} x_n \nu_n e^{-2 \pi inf} \ .\eqno (8)
$$
We define the spectral window, $V(f)$, to be the Fourier transform of $\nu_n$:

$$
V(f) = \sum_{n=-\Ntw}^{\Ntw} \nu_n e^{-2 \pi inf} \ ,
\ \ {\nu}_n = \int_{-1/2}^{1/2} V(f) e^{2 \pi inf}df . \eqno (9)
$$
To reduce the bias from spectral leakage, the taper should
be localized in the frequency domain. 
The expectation of the second moment of the tapered estimator is
$$
{\bf E} \left[  ({y}_{\nu}(f_1){y}^*_{\nu}(f_2)) \right]
=
\int_{-1/2}^{1/2} V(f-f_1) V^*(f-f_2)
S(f)df \ .\eqno (10)$$
We decompose the bias of the tapered spectral estimate into the  
narrow banded part from $|f'-f|<W$, and the broad-banded part.
For a taper to be useful in reducing the broad-banded bias, the sidelobes
of $|V(f'-f)|^2$ should decay faster than
$\left[ {\sin (N \pi (f-f^{\prime})) \over 
\sin ( \pi (f-f^{\prime}))}\right]^2 $.
Figure 1 plots the tapered and untapered kernels for $N=100$,
where $V(f)$ is the Slepian spectral window$^{21}$ 
which is described in Appendix A.
For $N\ge 100$, the sidelobes of both kernels are small on the linear scale.
If the spectrum varies rapidly on the logarithmic scale, these
sidelobes can cause considerable bias. Figure 2 displays the kernels
on the logarithmic scale for $N=300$.
Both the Fej\'{e}r kernel and the tapered kernel decay as 
$f^{-2}$, but the amplitude of the tapered sidelobes is reduced 
proportionally to $\exp(-NW)$ relative to the Fej\'{e}r kernel.

A third family of tapers which we use is the Tukey $\alpha$ 
split cosine taper where 
 
\medskip
$\nu_j = {1\over 2}[ 1 - \cos({\pi (j - 1/2) \over \alpha N}) ]$ \ \ \ \ \ \
for $1 \le j < \alpha N \ ;$

$\nu_j = 1$   \ \ \ \ \ \ \ \ \ \ \ \ \ \ \ \ \ \ \ \ \ \ \ \ \ \ \ \ \ 
for $ \alpha N \le j \le (1 - \alpha) N \ ;$

$\nu_j = {1\over 2}[ 1 - \cos({\pi (N- j + 1/2) \over \alpha N}) ] $ \ \ \
for $(1 - \alpha) N <j \le N \ .$
\medskip

The $\alpha$ parameter determines the extent of the tapering. As $\alpha$
increases, the bias protection improves at the cost of discarding more data.
Traditionally, $\alpha$ is set at .1, but recently Hurvich has shown that 
$\alpha N $ should be approximately constant as $N$ increases$^{26}$. 
Our RMSE comparison supports
Hurvich's conclusion, and we find that $\alpha N \sim $ 30--50 works best for
300-3000 point segments of the TFTR data.

We now restrict our consideration to {\it Gaussian stationary processes}.
In this case, the covariance of the quadratic tapered estimator is
$$
{\bf Cov} \  [|{y}_{\nu}(f_1)|^2 ,|{y}_{\nu}(f_2)|^2] =$$
$$
\left| \int_{-1/2}^{1/2} V(f-f_1) \Vbr(f-f_2)S(f)df \right|^2 
+
\left| \int_{-1/2}^{1/2} V(f-f_1) \Vbr(f+f_2)S(f)df 
\right|^2 
.\eqno (11)
$$
When $V$ is localized about zero, 
the second term  is only
important when both $f_1$ and $f_2$ are within a bandwidth of
zero frequency or the Nyquist frequency. We will usually neglect
this second term.
Generally, the Gaussian assumption is reasonable to estimate
${\bf Cov} \  [|{y}_{\nu}(f_1)|^2 ,|{y}_{\nu}(f_2)|^2] $ when $S(f)$
is smooth, but higher moments will be increasingly sensitive to
the Gaussian assumption. In Sec. V, we compare the Gaussian
error bars with a nonparametric estimate. We find that the  
Gaussian error bars actually overestimate the variance, probably due to
the coherent component of the signal.

We note that $|y_{\nu}(f)|^2$  is approximately 
distributed as a $\chi_2^2$ distribution,
and therefore has a variance almost exactly
equal to the square of its expectation.  {\it Thus
the quadratic taper estimate for the spectral density is
inconsistent, i.e. the variance of the estimate is fixed, and does 
not tend to zero as the number of data points, $N$, increases.}

Figure 3 plots the untapered periodogram estimate of the spectral density
for the entire 45,000 point data segment.
The rapid oscillation is characteristic
of the $\chi_2^2$ distribution of $|y(f)|^2$. 

When we use the entire 45,000 points of data, the bias of the periodogram
is small over much of the frequency range.
For nonstationary phenomena, a more typical data length is several hundred
data points. Therefore, we  examine the estimated spectrum for a typical
300 point data segment corresponding to a time interval of $.06$ millisec. 
Figure 4 plots the periodogram spectral estimate for the 300 point subsegment
beginning at time = 8.6 millisec. 
The spectrum is appreciably broader than the 45,000 point estimate of
Figs. 3 \& 5 due to the coarse resolution $\sim {1\over 300\Delta t} = 17 kHz$.
The coherence frequency scalelength for the random fluctuations 
to occur in the periodogram is also $\sim {1\over 300\Delta t}$.
{\it To reduce the variance of the spectral estimate, most statistical methods
average over a frequency interval or over time segments.}

\ \\
\ni
{\sc D. Consistent Spectral Estimators: the Smoothed Periodogram}
\ \\

The point tapered spectral estimate,
$\Shnu(f) = |\ynu(f)|^2$, has an approximately $\chtw$ distribution,
and therefore
a variance of $\Ebf[\Shnu]^2$ to leading order. Thus this raw spectral estimate
is statistically inconsistent in the sense that $\Shnu(f)$ does not
tend asymptotically to $S(f)$  as the time series length increases.
There are two standard remedies. 

First, if a number, $N_s$, of independent, statistically identical
time series are available, the spectral estimates from the individual series
may be averaged
to produce a mean estimate:
$\Sbr \equiv {1\over N_s}\sum_{k=1}^{N_s} \Sh_k{(f)}$. 
The mean estimate, $\Sbr$, has a variance approximately equal to $S(f)^2/N_s$.

Second, for a single time series, statistical consistency is normally achieved 
by smoothing the raw spectral estimate over a bandwidth, $W$.
This kernel smoothing decreases
the variance of the spectral estimate while increasing the bias.
The theory of optimal kernel smoothing is quite advanced$^{27,28}$. 
For our analysis, {\it we use a simple boxcar average with a fixed kernel
halfwidth.} 

The solid line in Fig.~5 plots the smoothed periodogram for the entire data
segment with a kernel halfwidth of 14 kHz.
The central peak at 1 MHz is partially coherent and is believed to
be due to reflection from the waveguide and vacuum vessel wall,
broadened by fluctuations at the plasma edge$^{16,17}$. 
{\it The secondary peak at 550 kHz is generated by fluctuations with phase
velocities in the electron drift  direction.} 
These fluctuations have a frequency spread of $\pm 100$ kHz.
The dashed line gives the smoothed tapered estimate using the same kernel
halfwidth of 14 kHz. We use a Tukey split cosine taper with $\alpha N = 100$.
Thus we trim only 200 data points out of 45,000. Nevertheless, a slight 
difference in the curves is visible at the highest frequencies. 
The dotted curve is the multitaper estimate, and will be discussed in the next
section.
 
The dashed line in Fig.~4 is the periodogram with $W = 70$ kHz. By smoothing,
we have reduced the variance at the cost of less frequency resolution.
The dotted curve is the smoothed tapered estimate using $W = 70$ kHz and
the Tukey split cosine taper with $\alpha N = 33$.
For $|f| > 1.6$ MHz, the smoothed periodogram is  
larger than the corresponding tapered curve.
Without tapering, the spectral estimates with small values of $S(f)$     
are artificially increased due to bias towards broadening due to the 
$\left[ {\sin (N \pi f) \over \sin ( \pi f)}\right]^2 $ kernel in Eq.~(6).

Figure 6 plots the smoothed tapered periodogram for the 300 point
subsegment of Fig.~4 for three different kernel halfwidths, 
40 kHz, 70 kHz, 120 kHz. As the kernel halfwidth increases, the spectrum
is smoothed and artificially broadened.

Figure 7 plots the relative RMSE
of the three different kernel halfwidths, averaged over 299 different
subsegments. The calculation of the RMSE is described in 
Sec. VI and Appendix D. The  RMSE increases with increasing bandwidth
near the spectral peaks. For larger frequencies, the spectral variation
is less and the optimal bandwidth is greater than 120kHz. Thus an
improved estimation procedure would be to use a small bandwidth near
the peaks and a larger bandwidth for the flatter parts of the spectrum.

\ \\

\medskip
\noindent
{\bf IV. Multiple Taper Spectral Analysis}
\vspace{.2in}

Multiple taper spectral estimation was presented in Ref. 3,
developed in Refs.~4-7, and has been used with good success 
in geophysical applications.
Thomson's theory treats spectral analysis as an inverse problem for the
integral equation of the first kind given in Eq.~(5). 
We solve the integral equation by
expanding in a set of eigenfunctions of a similar integral equation
with a band-limited kernel. 

As an alternative to the smoothed kernel estimate, Thomson$^3$ proposed to 
use a family of the orthogonal Slepian tapers$^{21}$, $\{ \nu^{(0)}_i \},
\{ \nu^{(1)}_i \}, \ldots, \{ \nu^{(K-1)}_i \}$, where 
$K \equiv 2NW $.
Appendix A describes the family of tapers which we use for both the
single and multiple taper analysis.
The bandwidth, $W$, is a free parameter for the Slepian tapers.
To begin the multitaper spectral analysis, we estimate each of
the $K$ tapered transforms:
$$y_{}^{(k)} (f) = \sum_{n=\Ntw}^{\Ntw} x_n \nu_n^{(k)} e^{-2 \pi inf} 
\ . \eqno (12) $$
In the frequency domain, this corresponds to the convolution equation:
$$\hat{y}^{(k)}_{\nu}(f_1) = \int_{-1/2}^{1/2}V^{(k)} (f-f_1) y(f)df 
= \int_{-1/2}^{1/2} V^{(k)}(f-f_1)dZ(f)\ \ .\eqno (13) $$
The different tapered estimates, $|\yk(f)|^2$, are  
statistically independent to 
$O\left(\left|\frac{W S'(f)}{S(f)}\right|^2\right)$.
Thus we have effectively created $2NW$ independent realizations of
a band-limited stochastic processes.
 
We then construct the $K$ raw estimates of the spectral density:
$\Sh_k(f) = |y^{(k)}|^2$, which are combined to produce a
``simple" multitaper spectral estimate: 
$$\Shb \equiv {1\over K}\sum_{k=0}^{K-1} \Sh^{(k)}(f) 
\ \ .\eqno (14) $$

In multiple taper estimation, each of {\it the estimates, $\yk(f)$,
is not only independent, but also centered at the frequency $f$.}
In contrast, smoothed single taper estimates are averages
of $y_{\nu}(f+f')$ as $f'$ varies, and $y_{\nu}(f+f')$ is centered
about the frequency $f+f'$, not $f$. As a result,
multiple taper analysis is more efficient, and has a lower bias
than the smoothed tapered periodogram.

Each of the eigentaper estimates, $\Sh_k(f)$, has a slightly
different bias and variance, and therefore, the  spectral estimate 
can be improved by replacing the simple average of the $\Shk$ with
a weighted linear combination of these estimates:
$$\Sh_c(f) \equiv \sum_{k=0}^{K-1} c_k(f) \Sh^{(k)}(f)
\ \ , \eqno (15) $$
\noindent
where  $c_k(f)$ are weights.  
The adaptive weightings differ from a uniform weighting, 
$c_k \equiv {1\over K}$, by terms of $O(S''(f_1)W^2)$ 
and of $O( 1- \lambda_k)$.
Appendix B presents three different adaptive weightings.
In this section, we use the {\it sequential deselection adaptive weighting}.
The main effect of the adaptive weightings is to downweight the last
tapers, ($k= K, K-1$), when $\sigma^2 (1 - \lambda_k) >> S(f)$.

The multitaper estimate of the entire data segment 
with the sequential deselection adaptive weighting is given as
the solid curve of Fig.~5.  We have used 252 tapers corresponding
to a bandwidth of 14 kHz. 
The spectral decay appears exponential  over four decades and not algebraic.
The adaptive multitaper estimate and the smoothed tapered periodogram
are virtually identical, showing that the difference in the estimates
tends to zero as $N$ increases.

Increasing the bandwidth and the corresponding number of tapers
decreases the variance and raises the smoothing bias.
Figure 8 compares the estimated spectrum for different values of $W$.
The spectral estimates in Figs. 6 \& 8 are similar and show that choosing 
the correct bandwidth is the most important aspect of spectral estimation.
The effective bandwidth of the smoothed tapered periodogram is larger due
to the finite support of the Fej\'{e}r kernel. Therefore, the optimal bandwidth
of multitaper analysis is larger than that of the
smoothed tapered periodogram. 




To compare the adaptive weighting of Eq.~(15) with  
the uniform weighting multitaper estimate of Eq.~(14), 
we compute both estimates on 300 point subsegments. (We average
over 299 different subsegments to reduce the variance of the estimate and
thereby emphasize the effect of the bias error.) Figure 9 plots the averages
of the two estimates.
Near 1 MHz, the two weightings are identical, while for $f > 1.5 $MHz,
the nonadaptive weighting oscillates with a frequency of $\frac{1}{N\Delta t}$.
The dashed line gives the uniform weighting multitaper estimate with the same
bandwidth, but without using the last two tapers, $k=K-1, K$. The artificial
oscillation is noticeably smaller, indicating that the last tapers are primarily
responsible. The variance of this estimate near 1 MHz is larger 
due to the absence of 
the last tapers. The adaptive multitaper estimate uses the maximum number of
tapers when the spectrum is large while effectively eliminating the most biased
tapers when the spectral density is much smaller than the average density.

In Sec. VI,
we also consider  ``$hybrid$'' methods where the multitaper estimate is 
then kernel smoothed. The total bandwidth is the multitaper bandwidth plus the
kernel smoother width. Hybrid methods offer the possibility of using
essentially all of the degrees of freedom in the frequency bandwidth (like
multitaper) with the flexibility and data  adaptivity of a $variable$
halfwidth kernel smoother$^{28}$.


\ \\
\noindent
{\bf V. Jackknife Estimates of Bias and Variance of Multiple Taper
Spectra}
\medskip

We now examine estimates of the variance of our multiple taper 
spectral estimate$^4$. Appendix C is a short review of the jackknife
resampling technique.
The most straightforward estimate of the variance of the multitaper 
spectral estimate is
$$
\sgh^2(f) = {1 \over K(K-1)} \sum_{k = 0}^{K-1} \left[ \
\hat{S}^{(k)} (f) - \bar{ \hat{S}}(f) \right]^2
\ , \eqno (16) $$
where $\bar{ \hat{S}}(f)$ is the simple arithmetic mean of the
multiple taper spectral estimate.
Eq.~(16) is adequate to estimate the variance of $\Shb(f)$. 
However, it tends to underestimate the probability of tail events
when the distribution is non-Gaussian, and it does not generalize
easily to the nonlinear adaptive weightings.

To robustify the variance estimate against non-Gaussian, large tail effects,
we transform to the logarithmic scale. 
To estimate $\ln(S)$, we use the logarithm of the mean:
$$
\widehat{{\rm ln}[\overline{S}(f)]}  \equiv
\ln \left( {1 \over K} \sum_{k = 0}^{K-1}  
\hat{S}^{({k})} (f) \right) \ \ ,\eqno (17) $$
instead of the mean of the logarithms:
$$
\overline{{\rm ln}[\hat{S}(f)]}  \equiv {1 \over K}
\sum_{k = 0}^{K-1}  {\rm ln}(
\hat{S}^{({k})} (f)) \ \ .\eqno (18) $$
Appendix C shows that Eq.~(18) has both larger bias and variance than Eq.~(17).
A related advantage of the logarithmic transformation is that
${\widehat{{\rm ln}S}}(f)$ converges in distribution more rapidly to
a Gaussian than does $\Sh(f)$.  Thus the confidence intervals for the
transformed variable are more accurate. 

Equation~(18) is biased downward,  and  the confidence intervals for
${\widehat{{\rm ln}S}}(f)$ are still based on the Gaussian assumption.
To correct for the downward bias and to  remove the Gaussian assumption,
we use the resampling technique of jackknifing. In accordance with Appendix C,
we use the delete-one samples to compute the variance as given by Eq.(C?).
The jackknife further reduces the effects of the large tail in the
probability  distribution, and thereby aids in the convergence of our
estimate in the variance of the combined estimate of $\ln(S(f))$.
The jackknife also allows us to propagate the effects of
the nonlinear adaptive weightings into the error analysis.


Figure 10 depicts the jackknife estimate of $2\sigma$ confidence interval for
$S(f)$. 
The dotted line gives the corresponding error estimates for 
$Gaussian$ processes.
The Gaussian error bars are actually larger 
than the empirical error bars near the spectral peaks. This unusual result
occurs because the spectral peaks are partially coherent i.e. contain 
a small number of waves which are oscillating in phase. In contrast, the
Gaussian error bar assumes not only that the process is Gaussian, but also
that it is zero mean (no coherent component), and it is this assumption that
probably breaks down.         

\ \\
\noindent
{\bf VI. Empirical Comparison of Spectral Estimation Methods}
\ \\

We have described the theoretical advantages of multitaper and
smoothed multitaper analysis
over the smoothed periodogram. The  extent of the advantages
is a function of the unknown spectrum and the sampling rate. We
now make a detailed study of the performance advantages of multitaper
analysis over conventional methods for the TFTR fluctuations
with 300-3000 point samples.  
Because the spectral range of the 
TFTR dataset is more than four orders of magnitude, we make our comparison
on the logarithmic scale.
We compute the estimated spectrum and the inferred RMSE
for the various methods on 299 subsegments of length 300 with 50 \% overlap. 
Because the true spectrum is unknown, we  use the 45,000 point
estimates in Fig.~5. The technical details of the comparison are given in 
Appendix D.

Figure 6 gives the relative RMSE, normalized to the converged spectrum, 
of the smoothed periodogram for different  
bandwidths. Clearly, the ``best'' bandwidth is a function of frequency.
To determine which bandwidth is optimal globally, we need to specify
how we wish to weight the RMSE as a function of frequency. We note that 
the errors are much larger at the spectral peaks. A simple averaging of the
RMSE or even the relative RMSE will be dominated by the fit to the 
spectral peaks. Instead, we concentrate on an accurate estimate of the
spectral density away from the peaks. Thus we  consider the relative RMSE
averaged over the frequency bands, $[200,900]$ kHz, $[1100,1900]$ kHz,
and $[2100, 2400]$ kHz.

The different methods have different effective 
bandwidths for the same ``official'' bandwidth, due to the finite support of
the kernel in Eq.~(10).
For each method separately,
we determine the optimal bandwidth by minimizing the relative RMSE 
averaged over the combined frequency bands, $[200,900]$ kHz
plus $[1100,1900]$ kHz plus $[2100, 2400]$ kHz.
Table 1 gives the integrated relative RMSE for each estimate. 
Table 2 gives
the kernel halfwidth which minimizes the integrated relative RMSE
for each method.

We begin by comparing the dependence of the relative RMSE on the taper 
shape. We then compare the three different adaptive weightings of
Appendix B. Finally, we conclude by comparing the hybrid, smoothed multitaper
with the best adaptive multitaper estimate and the best smoothed tapered 
periodogram. 

\ \\
{\sc A. Comparison of tapers for the smoothed periodogram}
\medskip

We compare the Tukey split cosine taper and the Slepian taper with no tapering
for the smoothed periodogram. Both the Tukey split cosine taper and the 
Slepian taper have auxilary parameters, $\alpha$ for the Tukey split cosine 
taper and $\wbr$ for the Slepian taper. In both cases, we optimize the taper
parameter with respect to the integrated relative RMSE. We find that
the best $\alpha N$ for the Tukey taper grows very slowly with $N$ 
(See Table 2). 
The integrated relative RMSE is a weak function of $\alpha N$.

Figure 11 plots the relative RMSE for each taper. The untapered 
periodogram performs so poorly that its relative  
RMSE is 2.5 times the RMSE
of the Tukey taper at high frequencies. For $|f| > 2.1$ MHz, the bias error
is most pronounced and the periodogram's  RMSE is $four$ times 
that of the tapered smoothed periodogram is used. 
As $N$ increases, the dominating
effect of nonlocal bias on the smoothed periodogram decreases. 

For $N=300$, the Tukey taper outperforms the Slepian taper at high frequencies,
and the Slepian taper does better at the spectral peaks. 
Since the Slepian taper windows more of the data
than does the Tukey taper, the Slepian taper offers more protection against
broad-banded bias at the cost of having a larger variance.
For $N=1000$, the Slepian taper loses much of its advantage near the peaks,
while continuing to have higher variance than the Tukey taper. 
As $N$ increases , the broad-banded  bias protection is less necessary.
By fixing $\alpha N \sim constant$,  a decreasing percentage of the data 
is tapered$^{26}$. In contrast, the Slepian taper cannot be relaxed towards no 
tapering and continues to taper a fixed percentage of the data even
when such strong bias protection is unnecessary.
Multitaper analysis circumvents this problem by  using  additional tapers.


\ \\
{\sc B. Comparison of adaptive weightings for  multitaper estimation}
\medskip

Figure 12 plots the normalized  RMSE for each  of the three adaptive
weightings of the multitaper estimate plus the unweighted estimate for $N=300$.
{\it Sequential deselection performs best, followed by the minimal loss 
weighting.} For $N = 300$, the sequential deselection  estimate has an average
relative error of 40 \%, which is appreciable. Table 1 shows that
our new adaptive weightings can reduce the error by 15 \%
relative to the error of the Thomson's adaptive weighting$^3$ (Eq. (B6))
and by 25 \% relative to the unweighted estimate. 

We have examined the contributions of the bias 
and variance to the total RMSE. 
Thomson's weighting has the lowest value 
of the bias. Our new weightings, sequential deselection and minimal loss,
deliberately bias the estimate downward to reduce the expected error (by
using the denominator $\frac{1}{K+1}$ instead of $\frac{1}{K}$). 
The minimal loss weighting has an additional weighting factor,
$\left({\lambda_k + {\sigma^2 \over S(f)} (1-\lambda_k)}\right)^{-1}$,
which is intended to correct the estimates for broad-banded bias. 
Unfortunately, our study finds that this correction almost always has
a larger bias error than does sequential deselection.
Presumably, the bias correction fails because the actual spectrum differs
significantly for the spectrum assumed in the model weighting. (See the 
second from the last paragraph in Appendix B.) 

When $N$ increases, 
the advantage of sequential deselection over previous multitaper methods
is reduced somewhat, but is still noticeable.
All four weightings produce similar estimates for $|f| < 1.5$ MHz,
however as the spectrum decreases and the bias increases, the  adaptive
weightings have noticeable differences.

\ \\ 
{\sc C. Multitaper\ --\ smoothed multitaper\ --\ 
smoothed periodogram comparison}
\medskip

Our final and most important result is that {\it the smoothed multitaper 
estimate outperforms the best ``pure'' multitaper and the best ``pure'' 
smoothed tapered periodogram by a moderate amount.}  
For short samples, $N \siml 300$, the smoothed four taper analysis reduces  
the relative RMSE by 14~\% in comparison  with the
adaptive multitaper and by 21~\% in comparison  with the
smoothed tapered periodogram. 
Figure 13 plots the normalized root mean square error, $RMSE(f) =
|\widehat{Var}(f)+ \hat{B}^2(f)|^{1/2} /  S_{Con}(f)$.
Figure 13 shows that the  smoothed multitaper dominates the other estimators
over the entire frequency range.
Figure 14 plots the ratios of the RMSEs: 
$\frac{RMSE(MT)}{RMSE(SMT)}$ and 
$\frac{RMSE(SP)}{RMSE(SMT)}$ as a function of frequency.
As $N$ increases, the advantage of smoothed multitaper decreases, but is
still apparent. For $N = 1000$, the smoothed five taper analysis reduces  
the relative RMSE by 7--8\% in comparison  with the
adaptive multitaper and smoothed tapered periodogram. 

We are uncertain as to exactly why the smoothed multitaper performs
better. One explanation is that without smoothing, the 
multitaper kernel,  $\sum_{k} |V_k(f)|^2 $,
has oscillatory sidelobes which becomes small at certain frequencies.
Thus the unsmoothed taper is not using these frequencies effectively
in the spectral estimate. By smoothing the multitaper estimate,
we are averaging the multitaper kernel and thereby using the degrees of
freedom in the frequencies where the unsmoothed sidelobes are very small.

For $N = 1000$, the sequential deselection  estimate has an average
relative error of 24 \%, which is effectively the same as the RMSE
for the smoothed periodogram with the Tukey taper. Thus the advantage of 
multitaper over the smoothed tapered periodogram is decreasing with $N$.
Multitaper analysis has more efficiency, because each of the taper estimates
$\Sh^{(k)}(f)$ is centered about the frequency of interest while the 
smoothed periodogram is essentially shifted spread an additional
half Rayleigh frequency, $\frac{1}{N\Delta t}$ on each side. Thus the
advantage of the multitaper will decrease with $\frac{1}{W N\Delta t}$,
which measures the number of Rayleigh frequencies in the bandwidth.


\newpage
\noindent
{\bf VII. Summary and Recommendations}
\medskip

We have compared  nonparametric estimators of the spectral density 
as measured by the RMSE error, normalized to the converged spectral
estimate, and  integrated over frequencies away from the spectral peaks
at 1 MHz and 2 MHz.  The smoothed four taper estimate 
outperforms  the ``pure'' multitaper and the  ``pure'' 
smoothed single taper periodogram. For $N=300$, 
the adaptive multitaper has a 16~\% larger RMSE, and the 
smoothed single taper periodogram has a  23~\% larger RMSE.
For $N=1000$ and $N=3000$, the smoothed multitaper estimate maintains
an advantage, but the difference is decreasing. 

Although we have only performed a detailed comparison on this particular 
dataset, we believe the relative ordering of the various methods is 
typical of data with smooth spectral densities which are at least partially
resolved ($\frac{1}{N\Delta t} <<$ characteristic frequency scalelength for spectral
variation). 
The performance ratios of the methods
depends critically on the amount of oversampling and on the spectral range.
{\it We therefore recommend that a smoothed multitaper estimate with four to
five different tapers be used in combination with the sequential deselection
adaptive weighting. When possible, the kernel smoother
should be  data adaptive such that the kernel halfwidth decreases when the
spectrum varies rapidly.}


Multitaper analysis uses all of the possible degrees of freedom in the 
bandwidth, $[f-W,f+W]$, while the smoothed tapered periodogram has 
its effective kernel support broadened  by at least $\frac{1}{2 N\Delta t}$
due to the spectral window support.
As a result, multitaper analysis has a smaller bias 
than the smoothed tapered periodogram for a given variance.
Since this additional broadening scales with the Rayleigh resolution,
the advantage of multitaper decreases as $N$ increases.   

Another popular technique for spectral estimation,
Welch's method$^{29}$,  divides the original time series
into $K$ separate pieces, and treats each segment as an 
independent realization.
{\it When the segments are not overlapped, Welch's method is an 
inefficient special case of multiple taper estimation.}
By artificially separating adjacent subsegments, information is destroyed.
Welch's method is inefficient because the first $K$ Slepian tapers are 
more strongly localized in frequency than the $K$ tapered subsegments.

All of the spectral estimation methods reduce
the variance of the estimate by effectively averaging over neighboring
frequencies. Hence each method has an optimal bandwidth which
minimizes the variance versus bias tradeoff. In our comparisons, 
the frequency bandwidths were chosen to 
minimize the integrated relative RMSE between the spectral estimate 
and the converged estimate using the 45,000 data point segment.

Another advantage of 
multitaper and smoothed multitaper analysis is that the variance is estimated 
nonparametrically using jackknife resampling. Since the TFTR microwave
fluctuations may be non-Gaussian in character,  the only reliable 
estimate of the error bars is given by resampling. Similarly, the sequential
deselection adaptive weighting is a nonparametric test to determine if
the bias is sufficiently large to delete the last taper. In our comparison,
sequential deselection outperformed the older adaptive weighting of Ref. 3
by up to 34~\%.

\ \\
\noindent
{\bf Acknowledgements}

We thank N. Bretz and R. Nazikian for many interesting discussions which 
stimulated this work and for detailed explanations of the physics of the 
TFTR microwave scattering experiment.
KSR thanks C. Hurvich for 
extensive discussions and advice on time series analysis,
and Alan Chave for providing his multitaper time series code. 
KSR \& AS  thank W. Sadowski and H. Weitzner for their support.
We thank the TFTR group for collecting and allowing us to analyze
the microwave scattering data. 
The referee's comments are gratefully acknowledged.
The work of KSR and AS was funded by the U.S. Department of Energy
Grant No. DE-FG02-86ER53223.


\ \\
\noindent
{\bf Appendix A: Maximally Frequency Localized Tapers}
\ \\

To minimize the bias error, associated with spectral leakage, we
seek basis functions/tapers which are concentrated about the particular
frequency of interest.
Given a bandwidth $W$, the Slepian functions$^{21}$, $V_k(f;N,W)$, are
defined  as the family of orthonormal solutions of
the extremal problem: Maximize
$${
\int_{-W}^{W} |V_k (f)|^2 df \over \int_{-1/2}^{1/2}
|V_k (f)|^2 df}
\ ,\eqno (A1)$$
subject to 
$$
V_k(f) = \sum_{n=-\Ntw}^{\Ntw} \nu^{(k)}_n e^{-2 \pi inf} .
\eqno (A2) $$
Equation A1 is most easily imposed
by substituting Eq.~(A2) directly into Eq.~(A1),
 and solving the variational problem in the time domain.
The first variation of Eq.~(A1) yields the linear eigenvalue problem:

$$ \sum_{n=-\Ntw}^{\Ntw}
A_{mn} (N,W)\nu_n^{(k)}  = \lambda_k \nu_n^{(k)} \ , \ \
 m =-\Ntw, \ldots \Ntw 
\ , \eqno (A3)$$
where the spectral concentration matrix, $\Ab(N,W)$, satisfies 
$$A_{mn} (N,W) \equiv \int_{-W}^{W} 
e^{2 \pi i(m-n)f} df= {\sin 2 \pi W(n-m) \over \pi (n-m)}.
\eqno (A4)$$
$\Ab(N,W)$ is a real symmetric matrix, and therefore, has
a complete set of $N$ orthonormal eigenvectors. 
In the Fourier domain, Eq.~(A3) corresponds to
$$
 \int_{-W}^W D_N (f-f^{\prime} )V_k (f^{\prime} )df^{\prime}
= \int_{-W}^W {\sin N \pi (f-f^{\prime}) \over \sin \pi (f-f^{\prime})}
V_k (f^{\prime}) = \lambda_k V_k (f)
\ . \eqno (A5)$$
The eigenvalue, $\lambda_k$, is the ratio of the integral of $|V_k(f)|^2$
in $[-W,W]$ to $[-{1\over2},{1\over2}]$,
and thus is always positive and bounded by one.
The  Slepian functions  are  known as 
prolate spheroidal wavefunctions. The discrete prolate spheroidal sequences
and the corresponding prolate spheroidal wavefunctions have a number
of special properties:

a) The Slepian functions are orthonormal on the interval, $[-1/2,1/2]$,
and orthogonal on the interval, $[-W,W]$: 
$$
\int_{-1/2}^{1/2} V_k(f)V_{k'}(f)df = \delta_{k,k'} \ , \ 
\int_{-W}^W V_k(f)V_{k'}(f)df =\lambda_k \delta_{k,k'}
\ . \eqno (A6)$$ 

b) The $k$th Slepian function has $k$ zeros in the interval $(-W,W)$.
Figure 15 displays the first four Slepian functions, $V_k(f;N,W)$,
for $N = 100$. 

c) $2NW$ eigenvalues, $\lambda_k$, are near one and the rest of the
eigenvalues are near zero. This corresponds to the $4NW$ degrees of freedom
in a band-limited signal of length $N$ when the center frequency is
not located within ${1\over N\Delta t}$ of zero. Table 3 displays
the eigenvalues versus the index $k$ for $N=300$.  

d) The sum of squares of the first $2NW$ Slepian functions,
$\sum_{k=1}^{2NW} |V_k(f;N,W)|^2$, converges to
the characteristic function of the interval $(-W,W)$ as $N$ increases:
$\sum_{k=1}^{2Nw} |V_k(f;N,W)|^2 \rightarrow \chi_{(-W,W)}(f).$ 
Figure 16 displays $\sum_{k=1}^{2Nw} |V_k(f;N,W)|^2 $.

e) The matrix,
$\Ab$, commutes with a tridiagonal matrix with well separated eigenvalues.
Thus the eigentapers, $\{ \nu^{(k)} \}$, may be determined numerically by
solving the eigenvalue equation of the tridiagonal matrix 
(see Appendix A in Ref. 6). As $N$ tends
to infinity, direct determination of the tapers through Eq.~(A3)
is exponentially ill-conditioned. For large taper lengths,
the only successful numerical method for determining the Slepian
eigentapers is to compute the eigenvectors of the commuting tridiagonal matrix.

\ \\
\ni
{\bf Appendix B: Adaptive Taper Weightings}
\ \\

In this appendix, we derive three adaptive weightings for the multiple
taper estimate -- sequential deselection, minimum expected error
weighting,
and Thomson's Wiener filter weighting. 
Our analysis assumes that the spectrum varies slowly on
the scale of the taper bandwidth $W$. Formally, we expand the properties
of the multitaper estimates in powers of $\frac{S^{''} (f) W^2}{S(f)}$. 
A generalization of Eq.~(12) shows that the spectral estimates from
two different orthogonal taper  are uncorrelated to 
$O\left(\left|\frac{W S'(f)}{S(f)}\right|^2\right)$. 

We decompose the bias of the $k$th spectral estimate into the  
narrow banded part from $|f'-f|<W$ and the broad-banded part.
To second order in $W$, we have
$$
\Ebf[\hat{S}^{(k)} (f)] = S(f)\lambda_k + S^{''} (f) D_{k} +
\dint |V_k (f-f')|^2 
S(f') df' \ \ , \eqno (B1)
$$
where $\dsint \equiv \int_{-1/2}^{-W} + \int^{1/2}_W$ and
$D_{k } = \int_{-w}^W {f'}^2 |V_k|^2 (f') df'$. 
From the Cauchy inequality applied directly to Eq.~(13), we have 
$$\int df \dint |V_k (f-f')|^2 S(f') df' \le \sigma^2 (1-\lambda_k) 
\ \ .\eqno (B2)$$
Thus we approximate the broad-band bias by replacing the local value
of the integral with  its average value, which yields the approximation:
$$
\Ebf [\hat{S}^{(k)} (f)] \sim S(f)\lambda_k + S'' (f) D_{k} +
\sigma^2 (1-\lambda_k) 
\ . \eqno (B3)$$
If $S''(f)$ is known or estimated, we can correct the $k$th spectral
estimate: $\hat{S}^{(k)}_{new} (f) =\hat{S}^{(k)}_{old} (f) - S'' (f) D_{k}$.
This estimator is unbiased to $O(W^4)$, and therefore, corresponds to 
higher order kernel estimators. In the absence of knowledge about
$S''(f)$, we will consider only the approximate bias given by  
$\sigma^2 (1-\lambda_k) $.

As Table 3 indicates, $1-\lambda_k$ is usually very small except for the 
last several tapers with $k\siml 2NW$. Therefore, a likely failure mode for
multitaper estimation is  that the last couple of spectral estimates,
$\Sh^{(k\sim 2NW)}$, are systematically larger than the other spectral 
estimates due to broad-banded bias. Our first adaptive weighting
is called sequential deselection and is designed to suppress this failure mode.

In {\it sequential deselection}, we begin by computing the mean and standard 
deviation of the first $K-1$ tapers.  We then require the last spectral 
estimate to satisfy the one sided F-test: $S^{(K)}(f) \le \Sbr(f) + 
\alpha_K\sgh(f)$.
If the $K$th estimate  fails the test, we delete it and repeat the test on
the next to last estimate. (We recompute $ \Sbr(f)$ and $\sgh(f)$ using only
the first $K-2$ tapers.) We modify this basic testing procedure in two ways. 
First, we require that no more than $20$ \% of the estimates are deleted.
Second, we recognize that $S^{(K)}(f)$  may pass the test due to random 
chance and that the $K-1$ estimate may be bad. Therefore, we automatically
test the $K-1$ estimate using the same one-sided test even when $S^{(K)}(f)$
passes the test. We require that two consecutive  spectral estimates pass
the one-sided test before terminating the testing procedure.
In practice, we find that only rarely are more than two estimates discarded.
Near the spectral peaks, the adaptive estimates use all of the estimates.
 
The second and third adaptive weightings assume that
the broad-banded bias is important and distributed evenly in the exterior
domain. Thus these adaptive weightings use the bias model: 
$E [\hat{S}^{(k)} (f)] \sim S(f)\lambda_k +\sigma^2 (1-\lambda_k)$.
The expected square error of the adaptive weighting of Eq.~(15) is
$$
\left[ \left( \sum_{k=0}^{K-1} c_k(f) \left(
S(f)\lambda_k + \sigma^2 (1-\lambda_k) \right)  - S(f) \right) \right]^2
 \ + \ \sum_{k=0}^{K-1} c_k(f)^2 \left(
S(f)\lambda_k + \sigma^2 (1-\lambda_k) \right)^2
\ , \eqno (B4)$$
where the first term is the square bias and the second term is the
Gaussian estimate of the variance.
Minimizing Eq.~(B4) with respect to $c_k$ yields
the {\it minimum expected error adaptive weighting}: 
$$
c_k(f) = {1 \over K+1} \left( { 1 \over
\lambda_k + {\sigma^2 \over S(f)} (1-\lambda_k)} \right)
\ . \eqno (B5)$$
Thus $c_k(f)$ downweights the last
tapers, ($k= K, K-1$), when $\sigma^2 (1 - \lambda_k) >> S(f)$.

In Refs.~3 \& 6, Thomson advocates a 
{\it Wiener filter  weighting} for $c_k$:
$$
g_k(f) = {1\over K} \left( {  \lambda_k \over
\left[\lambda_k + {\sigma^2 \over S(f)} (1-\lambda_k)\right]^2} \right)
\ , \ \ \ c_k(f) \equiv
 {K g_k(f) \over \sum_{k=0}^{K-1}{g_k}(f)} \  \ .\eqno (B6)$$
Thomson's $g_k(f)$ minimizes the expected error in estimating 
$dZ(f)$ instead of $S(f)$. The normalizing factor,
$K / \sum_{k=0}^{K-1}{g_k}(f)$, forces the adaptive weighting to
be unbiased when $S(f) \equiv \sigma^2$ at the cost of having a larger 
expected error. 
Equation (B6) downweights the last tapers more than the optimal 
weighting of Eq.~(B5). Thus the weighting of Eq.~(B6) is more
conservative  since it more strongly downweights the more questionable
$S^{(k)}$ and this gives it more robustness.

The second and third adaptive weightings essentially assume that
the spectrum is equal to $S(f)$ for $|f' -f|\le W$ and
$S(f')= {\sigma^2 -2wS(f)\over 1-2W}$ for $|f' -f|> W$.
This is a good assumption when the spectrum is roughly constant
over most of the domain and has local regions of much smaller amplitude.
Unfortunately, the more typical case is that the spectrum 
has local regions of much $larger$ amplitude. In this case, the
extent of the adaptive weighting should depend on the distance
to the spectral peaks. 
For the TFTR data at large
frequencies, $|f| > 1.7$ MHz, the broad-banded bias predominately comes from 
the remote spectral peak at 1 MHz. 
Thus the correction to the denominator of Eq.~(B5) should be 
$ {\sigma^2 \over S(f)}|V_k(f-1MHz)|^2$ instead of 
$ {\sigma^2 \over S(f)} (1-\lambda_k)$.  By using $(1-\lambda_k)$ instead of
$|V_k(f-1MHz)|^2$, we overcorrect for the broad-banded bias.
The same overcorrection is present in Thomson's (B6) weighting;
however, the normalizing factor reduces the bias at the cost of increasing the
variance by not downweighting the last tapers too much.

The minimum expected error adaptive weighting
is optimal only for the assumed spectrum.
The weak point of our analysis is that we treat $S(f)$ as given and not
estimated.
{\it In practice, the weights, $c_k(f)$, are computed iteratively
using the previous estimate of $S(f)$. 
Thus the total spectral estimate is nonlinear in $|y(f)|^2$.}

\newpage
\ni
{\bf Appendix C: Jackknife Estimates}
\ \\

The jackknife is a statistical ``resampling'' technique$^{18-20}$ to estimate 
the bias and variance
of complicated estimators, $\hat{\theta} = T_K (y_1 , \ldots , y_K )$.
For simplicity, we assume the $y_{\ell}$ are independent and identically
distributed, and that the estimator $\hat{\theta}_K$ is a symmetric function 
in its $K$ arguments.
The jackknife begins by estimating $\theta$ from the $K$ different delete-one
samples:
$$
\hat{\theta}_{(\bar{\ell})} \equiv T_{K-1} (y_1  , \ldots , y_{\ell -1} ,
y_{\ell+1} , \ldots , y_K ) \ \ .\eqno (C1)
$$

The jackknife reduces the bias by Aitken extrapolation. We assume
that $\hat{\theta}$ has a bias proportional to ${1\over K}$ and 
$\hat{\theta}_{(\bar{\ell})} $ has a bias proportional to ${1\over K-1}$. 
We define the $\ell$-th pseudovalue:
$$
p_{\ell} = K \hat{\theta} - (K-1) \hat{\theta}_{(\bar{\ell})} 
\ \ .\eqno (C2)$$
The bias proportional to $\frac{1}{K}$ has been eliminated 
from the pseudovalue. When the bias has a Taylor series expansion in 
$\frac{1}{K}$, the pseudovalues  have biases proportional to $\frac{1}{K^2}$.
The jackknife estimate of $\theta$ is the average of the $k$ pseudovalues
$$
\tilde{\theta} = {1 \over K} \sum_{i=1}^K p_i \ = \ 
K \hat{\theta}
- \ {K-1 \over K} \sum_{\ell =1}^K \hat{\theta}_{(\bar{\ell})}
\ \ .\eqno (C3)$$
The jackknife reduces the bias of the estimate, but in general raises the
variance of the estimate. The jackknife estimates the variance of
$\tilde{\theta}$ by assuming the pseudovalues, $p_{\ell}$, are
independent estimates of $\theta$:
$$
\hat{\sigma}_{jack}^2 = {\bf Var} \ [ \tilde{\theta} ] = {1 \over K(K-1)}
\sum_{\ell =1}^K (p_{\ell} - \tilde{\theta})^2 = {K-1 \over K}
\sum_{\ell =1}^K ( \hat{\theta}_{(\bar{\ell})} - \hat{\theta}_{(\bar{\cdot})}
)^2 \ \ ,\eqno (C4)$$
where
$$
\hat{\theta}_{(\bar{\cdot})} \equiv {1 \over K} \sum_{\ell =1}^K
\hat{\theta}_{(\bar{\ell})}
$$
is the arithmetic mean.

As $K$ increases, ${\tilde{\theta} - \theta \over \hat{\sigma}_{jack} }$ tends 
asymptotically to a $T$ distribution. The convergence is often slow, and
therefore ``variance  stabilizing'' transformations$^{20,30}$, such as the
logarithmic transformation for spectral estimation, are 
applied to the delete-one estimates $\hat{\theta}_{(\bar{\ell})} 
\rightarrow g(
\hat{\theta}_{(\bar{\ell})} )$ before jackknifing.

In particular, we wish to estimate $\ln(S(f))$ and the $variance$ of the 
estimate, given $K$ multitaper estimates. 
$\ln[\Sh^{(k)}(f)]$ has two sources of bias error. First, there is
bias because ${\bf \rm E}[\Sh^{(k)}(f)]$, as given by Eq.~(6), is not
equal to $S(f)$. For fixed $k$, this bias goes to zero as 
$N \rightarrow \infty$. Neglecting this effect, $\Sh^{(k)}(f)$ has a  
$\chi_2^2$ (or exponential) distribution, and so has its most probable value
at zero. As a result, the distribution of its logarithm has a 
very long lower tail. This lower tail causes a  second, more serious type 
of bias in the estimate of $\ln[\Sh^{(k)}(f)]$: 
$ {\bf E  [ }   \ln(\hat{S}^{(k)}) {\bf ]} =
\ln(S)- 0.577$
To reduce this bias, we average the $K$ estimates prior to taking the
logarithm: $\Shb \equiv {1\over K}\sum_{k=0}^{K-1} \Sh^{(k)}(f)$.
$\Shb$ has a $\chi_{2K}^2$ distribution and the expectation of its 
logarithm is given by
$$ 
{\bf E  [ }   \ln(\Shb) {\bf ]} =
\ln(S)+ B_{\chi} (K)
\ ,\eqno(C5)
$$
where the bias, $B_{\chi}(K)$, is given by
$$ 
B_{\chi} (K) =  \psi (K) \ -  \ln(K)
\ , \eqno(C6)$$
with $\psi$ being the digamma function. Thus the bias goes to zero as
$1 \over K$.
Similarly, the variance of $\ln(\Shb)$ is
$$ 
{\bf Var [}   \ln(\Shb) {\bf ]} = \psi'(K)
\ ,\eqno(C7)$$
where $\psi'$ is the trigamma function. 
Equation (18) computes the mean of the $\ln[\Sh^{(k)}(f)]$ rather than
the logarithm of the mean. Its variance is
$$ 
{\bf Var [}  \overline{{\rm ln}(\hat{S})} {\bf ]} = 
{1 \over K} 
\psi' (1) =   { \pi^2  \over 6K}
\ .\eqno(C8)$$
Table 4 shows that the variance of $\ln[\Shb]$ is lower than that of  
$\overline{{\rm ln}(\hat{S})}$. In Table 4, both quatities are given in
column 3 and 6 respectively. Column 2 gives the bias of $\ln[\Shb]$,
which compares favorably with $.577\over  K$, the bias of 
$\overline{{\rm ln}(\hat{S})}$.

Equations (C5)-(C8) are all based on the assumption that the stochastic
process is Gaussian. When the Gaussian  assumption is questionable,
the variance needs to be estimated empirically.
To minimize the effects of the long tail of $\ln(\hat{S}^{(k)})$, we use 
the jackknife estimate. For the log-spectral estimate, Eq.~(C1) becomes
$$ \hat{\theta}_{(\bar{\ell})} =
\ln[\Sh_{(\bar{\ell})} (f) ]= 
\ln  \left[  { 1 \over { K-1 } }  \sum^{K-1}_{k=0,k\ne \ell} 
\Sh^{(k)} ( f )  \right]
\ .\eqno(C9)$$
The jackknife variance become
$$ 
\hat{\sigma}_J^2 
= {K-1 \over K}  \sum_{\ell=0}^{K-1}  \left[  \ln[\Sh_{(\bar{\ell})} (f) ]
-  \ln[\SD (f)]   \right]^2
\ ,\eqno(C10)$$
where
$$ 
\ln (\SD ) = { 1 \over K } \sum_{\ell=0}^{K-1}  \ln[\Sh_{(\bar{\ell})} (f) ]
\ .\eqno(C11)$$
When the process is Gaussian, the expectation of jackknife variance estimate
is 
$$ 
{\bf E} [ \hat{\sigma}_J^2 ] = \
 {( K -1)^2 \over K } \left[ { 2 \over ( K - 2 )^2}
+ { 1 \over 2 }  \left[ \psi'({K-1\over 2})
- \psi'({K-2 \over 2} ) \right]  \right]
\ .\eqno(C12)$$
The derivation of Eq.~(C12) is complicated and
will be presented elsewhere. 
Asymptotically, Eq.~(C12) reduces to
$${\bf  E}[ \hat{\sigma}_J^2  {\bf ]} \sim
{ (K-1)^2 (K - 3)  \over  K (K-2)^3}  
\ .\eqno(C13)
$$
In Table 4, Column  4 gives the expectation of the jackknife variance 
estimate, $\sigma^2_J$, and column  5 gives its asymptotic form for large $K$.
Thus, the jackknife variance estimate slightly overestimates the actual
variance, $\psi'(K)$.
Column 7 gives the ratio of column 6 to column 4 and shows the superiority
of the jackknife estimate.

\newpage
\ni
{\bf Appendix D: Empirical estimation of the expected  error} 
\ \\

The average square error consists of two pieces, the variance and the 
square bias. The true spectrum is unknown, and therefore, we estimate it
using the entire 45,000 point segment with a bandwidth of 14 kHz,
corresponding to 500 degrees of freedom. Figure 5  illustrates that
the multitaper estimate and the smoothed periodogram are nearly identical. 
We will treat the adaptive multitaper estimate using sequential deselection
as the converged value of the true spectrum, $S_{Con}(f)$. 
To quantify the difference, 
we compute the difference of the two estimates in Fig.~5 
divided by the multitaper estimate, $
{S_{SP}(f) \over S_{MT}(f)} - 1$,
where the subscripts, $MT$ and $SP$, stand for multitaper and smoothed
Tukey tapered periodogram respectively. We find that on average
the two estimates seldom differ by less than 1.1 percent. 

We estimate the bias of the various methods by
$$\hat{B}(f) = {1\over N_s} \sum_{i=1}^{N_s}
\Sh_{i}(f) - S_{Con}(f)
\ . \eqno (D1)$$
Our estimate of the bias neglects the error in $S_{Con}(f)$ and assumes  
that the spectrum is stationary. We expect that the square bias
is proportional to $W^4$ and that the variance is inversely proportional to
the number of degrees of freedom. Thus the error in the 300 point estimates
is  much larger than that of the 45,000 point estimate. The assumption
of stationarity is less valid. Figure 17 plots the multitaper estimates with
$W = 20$ kHz for the first third, the second third and the final third of the
data segment. Differences are visible only at the secondary maximum at 
550 kHz $\pm$ 100 kHz. Physically, this is interesting because this secondary
maximum represents the dominant drift wave frequencies.
Thus the electron drift wave fluctuation level is growing on the ten 
millisecond time scale. In Ref. 11, we explore nonstationary 
plasma fluctuations. 
For the purpose of this article, the examination of stationary 
spectra, our empirical convergence study may not be relevant in the
400--600 kHz frequency range. 

To reduce the influence of nonstationarity on our estimate of the variance,
we compare $\Sh_{i}(f)$ with the average of $\Sh_{i-1}(f)$ 
and $\Sh_{i+1}(f)$: 
$$\widehat{Var}(f) = {2\over 3(N_s-2)} \sum_{i=2}^{N_s-1}
\left(\Sh_{i}(f) - { \Sh_{i-1}(f) +\Sh_{i+1}(f) 
\over 2} \right)^2 
\ , \eqno (D2)$$
where we use nonoverlapping subsegments to compute the variance. 

\ \\

\np
{\bf Bibliography}{}
\begin{enumerate}

\item A.J. Wooten, B.A. Carreras, H. Matsumoto, 
K. McGuire, W.A. Peebles, Ch.P. Ritz, P.W. Terry, S.J. Zweben, 
Phys. Fluids B, {\bf 2},  2879 (1990).
\item P. Liewer, Nucl. Fusion {\bf 25},  543 (1985).



\item{D.J. Thomson, 
{Spectrum estimation and harmonic analysis.}
{\   Proc. I.E.E.E.}, {\bf 70}, 1055 
 (1982).}

\item
{D.J. Thomson and A.D. Chave,  
in {\it Advances in spectrum analysis},
edited by S. Haykin, (Prentice-Hall, New York 1990) 
Ch. 2,  pg. 58-113.}

\item
{C.T. Mullis  and  L.L. Scharf,   
in {\it Advances in spectrum analysis},
edited by S. Haykin, (Prentice-Hall, New York 1990) 
Ch. 1, {pg. 1-57}.}

\item
{D.J. Thomson, %
{\   Phil. Trans. R. Soc. Lond. A}, {\bf 332}, 539 (1990).} 

\item
{J. Park, C.R. Lindberg, and F.L. Vernon,   
{\   J. Geophys. Res.}, {\bf 92B}, 12,675 
 (1987).}

\item{M.B. Priestley,  
{\it Spectral analysis and time series.} 
({Academic Press}, New York 1981).  }

\item{L.H. Koopmans,  
{\it The spectral analysis of time series.} Ch. 8, 
({Academic Press}, New York 1974).  }

\item 
{U. Grenander and M. Rosenblatt,  
{\it Statistical analysis of stationary time series.} 
({Wiley}, New York 1957).  }

\item K.S. Riedel,  A. Sidorenko, N. Bretz, 
D.J. Thomson,
Spectral  density estimation of plasma 
fluctuations II: Nonstationary analysis  of ELM spectra. 
Published in this issue, Physics of Plasmas {\bf 1} page ? (1994).

\item C.P. Ritz, E.J. Powers and R.D. Bengtson, Phys. Fluids B {\bf 1},
 153 (1989).

\item D.J. Grove and D.M. Meade,
Nuclear Fusion {\bf 25},  1167 (1985).

\item N. Bretz, P. Efthimion, J. Doane, and A. Kritz, Rev. Sci. Instr.
{\bf 59},  1538 (1988).

\item N. Bretz, R. Nazikian, W. Bergin, M. Diesso, J. Felt, M. McCarthy, 
Rev. Sci. Instr. {\bf 61}, 3031 (1990).

\item N. Bretz, R. Nazikian, and  K. Wong, {\it Proceedings  of the
17th European Phys.  Soc. Conf.}, 
(European Phys.  Soc., Amsterdam, 1990) p. 1544.

\item  R.~Nazikian, N.~Bretz,  E.~Fredrickson, 
Y.~Nagayama, E.~Mazzucato, K. McGuire, H.K.~Park, G.~Taylor, 
A.~Cavallo, M.~Diesso, J.~Felt,
{\it Proceedings  of the
18th European Phys.~Soc.~Conf.}, 
(European Phys.~Soc., Berlin  1991) Vol. I p. 265.


\item{R. Miller, 
{\   Biometrika}, {\bf 61},  1 
(1974). }

\item{C.F.J.  Wu, 
{\   Ann. Stat.}, {\bf 14},  1261 
 (1986). }

\item{N. Cressie, 
{\   J. Roy. Stat. Soc. B}, {\bf 43},  177 
 (1981). }

\item
{D. Slepian, 
{\   Bell Syst. Tech. J.}, {\bf 57}, 1371 
(1978).}

\item J. Sheffield, {\it Plasma Scattering of Electromagnetic Radiations.}
(Academic Press, New York, 1975).

\item N. Bretz, Plasma Phys. {\bf 38},  279 (1987).

\item T.P. Hughes and S.R.P. Smith, Plasma Phys. {\bf 42},  215 (1989).

\item Y.J. Kim, K.W. Gentle, Ch. P.Ritz, T.L. Rhodes, R.D. Bengston, 
Phys. Fluids B {\bf 3} 674 (1991). 

\item{C. Hurvich,  
{\   Biometrika}, {\bf 75},  485} (1988).

\item
W.  Haerdle, 
{\it Applied nonparametric regression.} 
(Cambridge University Press, Cambridge, New York, 1990).

\item{H.G. Mueller, and U. Stadtmueller, 
{\   Annals of Statistics}, {\bf 15}, 182
 (1987). }

\item{P.D. Welch,
{I.B.M J. Research Devel.} {\bf 5}, 141  (1961).}

\item{R.G.Miller, 
{\ Ann. Math. Statist.}, {\bf 61},  567 
(1968). }


\end{enumerate}

\newpage

\begin{center}

\begin{tabular}{|l|l|l|}\hline
 $k$  & $\;\;\;\;\;\;\;\;\;\;\lambda_k$ & $\;\;\;\;1-\lambda_k$
\\ \hline 
0 & 0.99999999999996 & 3.62$\cdot 10^{-14}$ \\
1 & 0.999999999995 & 4.65$\cdot 10^{-12}$ \\
2 & 0.9999999997 & 2.90$\cdot 10^{-10}$ \\
3 & 0.999999989 & 1.14$\cdot 10^{-8}$ \\
4 & 0.9999997 & 3.18$\cdot 10^{-7}$ \\
5 & 0.999993 & 6.60$\cdot 10^{-6}$ \\
6 & 0.99989 & 1.05$\cdot 10^{-4}$ \\
7 & 0.9987 & 1.31$\cdot 10^{-3}$ \\
8 & 0.9875 & 1.25$\cdot 10^{-2}$ \\
9 & 0.9157 & 8.43$\cdot 10^{-2}$ \\
\hline
\end{tabular}

\end{center}
\ \\

\  \\
\centerline{ Table 3: Eigenvalues for $N=300, W =91$ kHz}

\  \\

\newpage

\begin{center}

\begin{tabular}{|l|l|l|l|l|l|l|}\hline
 $K$  & Bias$_{\chi}(K)$ & $\psi'{(K)}$& Eq. (C12) & Eq.(C13)& Eq. (C8)& 
${Eq.~C8 \over Eq.~C12}$
\\ \hline 
 
3 &    0.17582 & 0.39493 & 0.47342 & 0.00000 & 0.54831 & 1.15818 \\	
4 &    0.13017 & 0.28382 & 0.32610 & 0.28125 & 0.41123 & 1.26105 \\
5 &    0.10332 & 0.22132 & 0.24732 & 0.23703 & 0.32898 & 1.33019 \\
6 &    0.08564 & 0.18132 & 0.19879 & 0.19531 & 0.27415 & 1.37905 \\	
7 &    0.07312 & 0.15354 & 0.16605 & 0.16457 & 0.23499 & 1.41515 \\
8 &    0.06380 & 0.13313 & 0.14251 & 0.14178 & 0.20561 & 1.44279 \\
9 &    0.05658 & 0.11751 & 0.12479 & 0.12439 & 0.18277 & 1.46459 \\
10 &   0.05083 & 0.10516 & 0.11097 & 0.11074 & 0.16449 & 1.48220 \\
11 &   0.04614 & 0.09516 & 0.09991 & 0.09976 & 0.14953 & 1.49671 \\
12 &   0.04224 & 0.08690 & 0.09084 & 0.09075 & 0.13707 & 1.50887 \\
13 &   0.03895 & 0.07995 & 0.08329 & 0.08322 & 0.12653 & 1.51919 \\
14 &   0.03613 & 0.07404 & 0.07689 & 0.07684 & 0.11749 & 1.52807 \\
15 &   0.03370 & 0.06893 & 0.07140 & 0.07137 & 0.10966 & 1.53578 \\
16 &   0.03157 & 0.06449 & 0.06664 & 0.06662 & 0.10280 & 1.54254 \\
17 &   0.02970 & 0.06058 & 0.06248 & 0.06246 & 0.09676 & 1.54852 \\
18 &   0.02803 & 0.05712 & 0.05881 & 0.05879 & 0.09138 & 1.55384 \\
19 &   0.02654 & 0.05404 & 0.05554 & 0.05553 & 0.08657 & 1.55860 \\
20 &   0.02520 & 0.05127 & 0.05262 & 0.05261 & 0.08224 & 1.56290 \\
\hline
\end{tabular}

\end{center}
\ \\

\  \\
\centerline{ Table 4: Jackknife Log-spectral Estimates}

Column 	1 gives $K$, the number of tapers in the spectral estimate.
Column 2 gives 	the bias of $ \ln(\Shb)$ for $K$ tapers and  
column  3 gives its variance.
Column  4 gives the expectation of the jackknife variance 
estimate, $\sigma^2_J$, and column  5 gives its asymptotic form for large $K$.
Column 6 gives the variance of the estimate of Eq.~(18).
Column 7 gives the ratio of column 6 to column 4 and shows the superiority
of the jackknife estimate.

\  \\

\includepdf[pages=-,pagecommand={}]{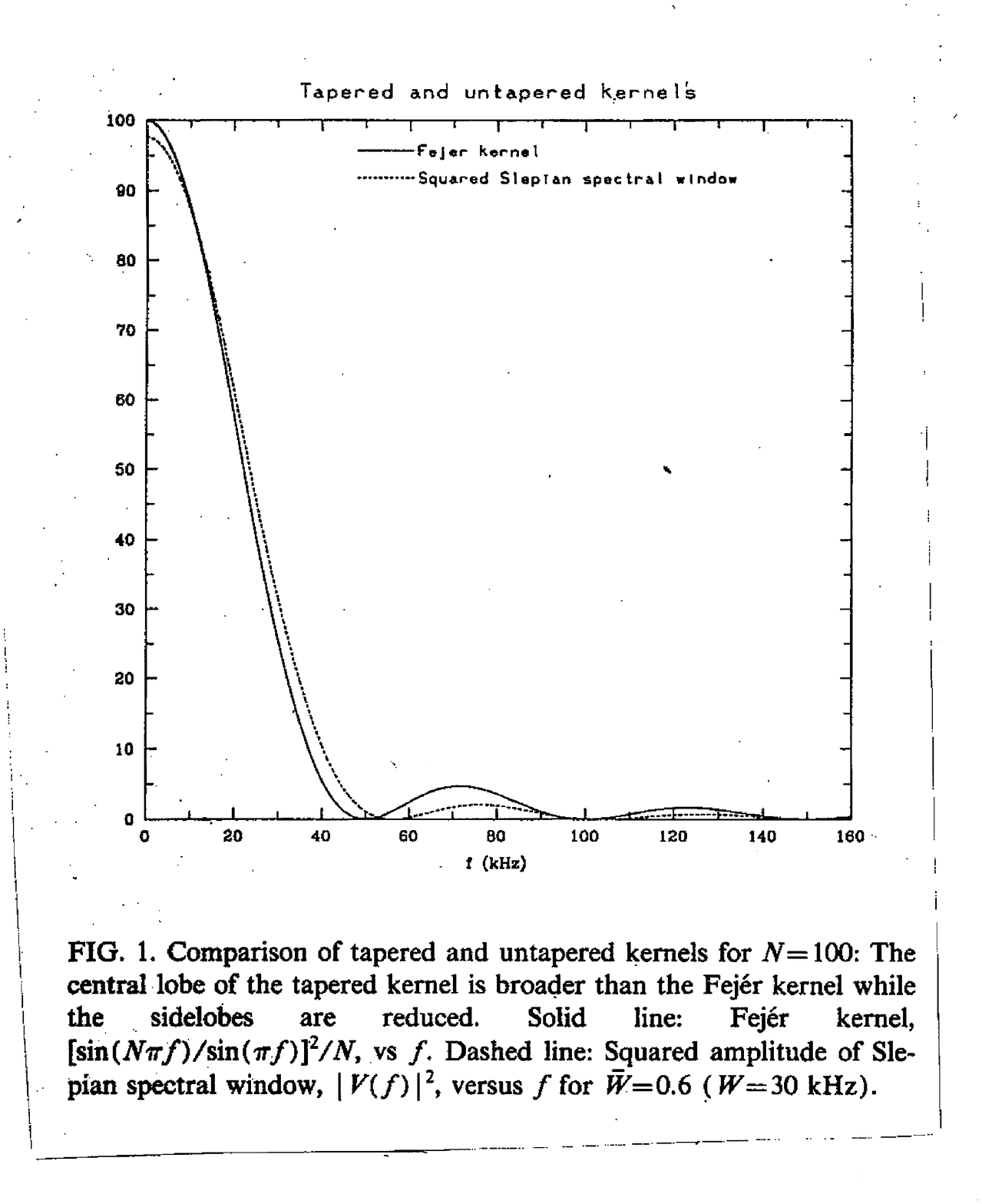}
\newpage
{{  Figure Captions:}
\vspace{.2in}



Figure 1: Comparison of tapered and untapered kernels for $N=100$:
The central lobe of the tapered kernel is broader than the Fej\'{e}r kernel 
while the sidelobes are reduced.
 
Solid line: Fej\'{e}r kernel,
$\frac{1}{N}\left[ {\sin (N \pi f) \over \sin ( \pi f)}\right]^2 $, 
versus $f$. 

Dashed line: 
Squared amplitude of Slepian spectral window, $|V(f)|^2$, 
versus $f$ for $\wbr = .6$ ($W = 30$ kHz).

\vspace{.2in}

Figure 2:  Fej\'{e}r kernel and squared Slepian spectral window $\wbr = .6$
($W = 10$ kHz), 
log scale  for $N=300$. Both the tapered and untapered kernels decay as 
$f^{-2}$. The amplitude of the tapered sidelobes is reduced 
proportional to $\exp(-NW)$ relative to the Fej\'{e}r kernel.

\vspace{.2in}

Figure 3: Periodogram of entire 45,000 point segment, log scale.
The rapid oscillations occur because the point estimates, $|y(f)|^2$,
are nearly uncorrelated at frequencies of $1/N\Delta t$ apart.
If the fluctuations were Gaussian and resolved in frequency, the periodogram 
would have a $\chi_2^2$ distribution. 

\vspace{.2in}

Figure 4: Spectral estimates of 300 point segment beginning at $t= 8.6$, 
log scale. The high variance of the unsmoothed periodogram obscurs the 
systematic differences due to tapering.

Solid line: Periodogram.

Dashed line: Smoothed periodogram with $W = 70$ kHz.

Dotted line: Smoothed tapered periodogram with $W = 70$ kHz
using the Tukey split cosine taper with $\alpha N =33$.


\vspace{.2in}

Figure 5: Smoothed spectrum of entire 45,000 point segment, $W =14$ kHz. 
The central peak at 1 MHz is partially coherent and is believed to
be due to fluctuations at the plasma edge.
The secondary peak
at 550 kHz is generated by fluctuations which are propagating in the
electron drift direction. 
These fluctuations have a frequency spread of $\pm 100$ kHz.

Solid line: Multitaper estimate with sequentual deselection.

Dashed line: Tukey split cosine taper with $\alpha N = 100$.

Dotted line: Periodogram.


\vspace{.2in}


Figure 6: Smoothed tapered periodogram for the 300 point
subsegment of Fig.~4 for three different kernel halfwidths, 
40 kHz, 70 kHz, 120 kHz. As the kernel halfwidth increases, the spectrum
is smoothed and artificially broadened. The area under the 1 MHz 
peak is approximately conserved.

Solid line: $W = 40$ kHz.

Dashed line: $W = 70$ kHz.

Dotted line: $W = 120$ kHz.

\vspace{.2in}

Figure 7: Relative RMSE
of the three different kernel halfwidths, averaged over 299 different
subsegments. The calculation of the RMSE is described in 
Appendix D. $W = 40$ kHz does best at the spectral peaks, $W = 120$ kHz
does best at the high frequency, and $W = 70$ kHz does best  overall.

Solid line: $W = 40$ kHz.

Dashed line: $W = 70$ kHz.

Dotted line: $W = 120$ kHz.

\vspace{.2in}

Figure 8: Adaptive multiple taper spectral estimate of 300 point segment 
beginning at $t= 8.6$. 
Figs.~6 \& 8 are similar, showing that choosing 
the correct bandwidth is the most important aspect of spectral estimation.
For the same ``official'' bandwidth,
the effective bandwidth of the smoothed tapered periodogram
is larger than that of the multitaper estimate. 
This additional broadening
is proportional to the Rayleigh resolution. 
For $N=300$, the Rayleigh resolution frequency, 
$\frac{1}{300 \Delta t } \siml 17$ kHz, is a significant fraction of
the  spectral estimation halfwidth.
Thus we use  slightly larger bandwidths for the multitaper estimate.

Solid line: $W = 58$ kHz.

Dashed line: $W = 91$ kHz.

Dotted line: $W = 141$ kHz.

\vspace{.2in}

Figure 9: Multiple taper spectral estimate averaged over 
299 overlapping 300 point segments. The adaptive weightings downweight
the last several tapers when the inferred broad-banded bias is too large.  

Solid  Line: Uniform weighting of $\frac{1}{K+1}$.

Dashed line: Uniform weighting without the last two tapers.

Dotted line: Adaptive multitaper using sequential deselection.



\vspace{.2in}

Figure 10: Resampled versus Gaussian $2\sigma$ confidence interval comparison:
The jackknife error bars are calculated by resampling 
the empirical distribution of
the individual multitaper estimates, $S^{(k)}(f)$. Thus the jackknife
error bars correspond to the actual distribution of the random process.
The dotted line gives the error bars for $Gaussian$ processes,
calculated from the fourth moment identity of Gaussian processes. 
The Gaussian error bars are actually larger 
than the empirical error bars near the spectral peaks. This prrobably
occurs because the spectral peaks are partially coherent while
Gaussian error bars assumes  that there is no coherent component.


Solid line: Jackknife error bars for 300 points. 

Dotted line: Gaussian error bars.


\vspace{.2in}

Figure 11: Normalized RMSE: $RMSE(f) /  S_{Con}(f)$ of various 
tapers for the smoothed periodogram,
where $RMSE(f) =|\widehat{Var}(f)+ \hat{B}^2(f)|^{1/2}$.
The RMSE is calculated from 299 estimates of 300 point overlapping
subsegments.

Solid  line: Slepian taper, $W = 75$ kHz, $\wbr  = 1.0$.

Dashed line:  Tukey split cosine taper, $W = 70 $ kHz, $\alpha N  = 33$.

Dotted line: No tapering, $W = 65$ kHz.

\vspace{.2in}

Figure 12: Normalized RMSE: $RMSE(f) /  S_{Con}(f)$ of various 
adaptive multitaper weightings, 
where $RMSE(f) =|\widehat{Var}(f)+ \hat{B}^2(f)|^{1/2}$.

Solid  line: Wiener noise adaptive weighting (Eq. B6), $W = 83$ kHz.

Dashed line: Minimal expected loss adaptive weighting (Eq. B5), $W = 91$ kHz.

Dotted line: Adaptive multitaper using sequential deselection, $W = 91$ kHz.

Dashed--Dotted line: Uniform weighting, $W = 83$ kHz.

\vspace{.2in}

Figure 13: Normalized RMSE for multitaper\ --\ smoothed multitaper
\ --\ smoothed periodogram comparison. {\it The smoothed four taper estimate 
outperforms either ``pure'' method.} 
The RMSE is calculated from 150 estimates of 300 point nonoverlapping
subsegments.

Solid  line: Smoothed 4 taper hybrid estimate, $W = 60$ kHz, $\wbr = 2.46$.

Dashed line: Smoothed tapered periodogram with $W = 70$ kHz
using the Tukey split cosine taper with $\alpha N =33$.

Dotted line: Adaptive multitaper using sequential deselection, $W =91$ kHz. 

\vspace{.2in}

Figure 14: Performance ratios of multitaper and smoothed tapered periodogram
to the smoothed multitaper.


Dashed line: RMSE of the smoothed tapered periodogram 
divided by RMSE of the smoothed multitaper.

Dotted line: RMSE of the adaptive multitaper
divided by RMSE of the smoothed multitaper. 

\vspace{.2in}

Figure 15: First four Slepian spectral windows for $N=100$, $W =100 kHz$.
The zeroth order taper resembles a standard windowing function.
Each  successive window has an additional oscillation.
Thus the spectral windows resemble Hermite polynomials modulated by a Gaussian.

\vspace{.2in}

Figure 16: The effective kernel shape, $\sum_{k=1}^{K=2NW} |V_k(f;N,W)|^2 $,
versus frequency. As $N$ increases for fixed $W$, the kernel becomes
a step function with bandwidth, $W$.



\vspace{.2in}

Figure 17: Adaptive multitaper spectral estimates  of three nonoverlapping 
15,000 point segments using sequential deselection $W = 20$ kHz, $120$ tapers.
Differences are visible only at the secondary maximum at 
550 kHz $\pm$ 100 kHz. The secondary
maximum represents the electron drift wave fluctuations, which are growing 
on the ten millisecond time scale. Due to this nonstationarity,
our empirical convergence study may not be relevant in the
400--600 kHz frequency range. 

Solid  Line: $t = 4.1 - 7.1$ millisec.

Dashed line: $t = 7.1 - 10.1$ millisec.

Dotted line: $t = 10.1 -  13.1$ millisec.

\end{document}